\documentclass[aps,twocolumn,showpacs,preprintnumbers,nofootinbib,floatfix]{revtex4}
\usepackage{graphicx}
\usepackage{epsfig}
\usepackage{dcolumn}
\usepackage{subfigure}
\usepackage{multirow}
\usepackage{amsmath}

\def\be{\begin{equation}}
\def\ee{\end{equation}}
\def\bea{\begin{eqnarray}}
\def\eea{\end{eqnarray}}

\def\d#1#2{\frac{\displaystyle #1}{\displaystyle #2}}
\def\no{\nonumber}
\def\p{\partial}
\newcommand{\omits}[1]{}

\def\bsp{\be\begin{split}}

\def\bes{\be  \begin{split}}

\def\p{\partial}

\def\eps{\epsilon}

\newcommand{\Rmnum}[1]{\expandafter\@slowromancap\romannumeral #1@}

\def\PRD{{Phys. Rev. D}}
\def\PRL{{Phys. Rev. Lett. }}

\def\PLB{{Phys. Lett. B}}

\def\CQG{{Class. Quant. Grav. }}

\def\IJTP{{Int. J. Theor. Phys. }}

\def\JHEP{{JHEP}}

\begin{document}

\title{Phase transition and thermodynamic stability of topological black holes in Ho\v{r}ava-Lifshitz gravity}
\author{Meng-Sen Ma$^{a,b}$\footnote{Email: mengsenma@gmail.com; ms\_ma@sxdtdx.edu.cn}, Ren Zhao$^{a,b}$, Yan-Song Liu$^{a}$}

\medskip

\affiliation{\footnotesize$^a$Department of Physics, Shanxi Datong
University,  Datong 037009, China\\
\footnotesize$^b$Institute of Theoretical Physics, Shanxi Datong
University, Datong 037009, China}

\begin{abstract}
On the basis of horizon thermodynamics, we study the thermodynamic stability and $P-V$ criticality of topological black holes constructed in Ho\v{r}ava-Lifshitz (HL) gravity without the detailed-balance condition (with general $\epsilon$). In the framework of horizon thermodynamics, we do not need the concrete black hole solution (the metric function) and the concrete matter fields. It is shown that the HL black hole for $k=0$ is always thermodynamically stable. For $k=1$, the thermodynamic behaviors and $P-V$ criticality of the HL black hole are similar to those of RN-AdS black hole for some $\eps$. For $k=-1$, the temperature is classified into six types by their different features. Among them, we mainly focus on the type with a triply degenerate thermodynamic state. It is also shown that there is a `` thermodynamic singularity" for the $k=-1$ HL black hole, where the temperature and Gibbs free energy both diverge apart from a special pressure $P_s$.

\end{abstract}

 \maketitle

\section{Introduction}

Since the discovery of Hawking radiation, black hole thermodynamics has attracted a lot of interest. It shows that gravitational systems like black holes can also be
thermodynamic systems. Not only that, the study of black hole thermodynamics may shed light on the understanding of the theory of quantum gravity.
A consistent theory of quantum gravity should succeed in addressing the thermodynamic nature of black holes. Black holes do not only have the standard thermodynamic
quantities, such as temperature and entropy, but also possess
abundant phase structures like the Hawking-Page phase
transition \cite{HP} and critical phenomena similar to those
in ordinary thermodynamic systems.
Especially, it is found that the charged AdS black hole may have the similar phase transition and critical behavior to that of Van der Waals liquid/gas system \cite{Chamblin,Lemos,Wu}.
There are also extensive studies on the thermodynamics and $P-V$ criticality of several kinds of black holes in an extended phase space by taking the cosmological constant as pressure \cite{MMC,Kastor,Dolan,Mann1,Mann2,Wu2,LYX,Ma1,Ma3,ZhaoHH.2015}.

In fact, there is another route to explore the relationship between a gravitational system and its relevant thermodynamic properties. It is the framework of horizon thermodynamics proposed by Padmanabhan \cite{Pa1,Pa2}.
It is shown that Einstein's equations for a spherically symmetric spacetime can be written in the form of the first law of thermodynamics: $dE=TdS-PdV$. This makes the connection between gravity and thermodynamics more closely. The pressure $P$ is the $(rr)$ component of energy-momentum tensor. In this case, only two pairs of thermodynamic variables exist, which are the intensive quantities $(T,~P)$ and the extensive quantities $(S,~V)$.
In this framework the thermodynamic properties are directly related to the gravitational theories under consideration.  The details of matter content are not important and the concrete black hole solutions are also not necessary. This approach has also been extended to the other theories of gravity, such as Lovelock gravity \cite{Pad:2006} and Ho\v{r}ava-Lifshitz theory \cite{Cai:2010}.
In this framework, we have studied the phase transitions and thermodynamic stabilities of black holes in general relativity and Gauss-Bonnet gravity \cite{Ma2}. Recently, in \cite{Hansen} the authors discussed the phase structures of Lovelock black holes in horizon thermodynamics and compared with that in extended phase space with variable cosmological constant.

Recently much attention has been focused on the Ho\v{r}ava-Lifshitz (HL) theory \cite{Horava}, which
is a power-counting renormalizable gravity theory and can be regarded as a UV complete candidate for general relativity. The static, spherically symmetric solutions in HL gravity have been obtained in \cite{Lu:2009, Cai:2009-PRD,Cai:2009-PLB,EK:2010}. The thermodynamic quantities of topological HL black holes are calculated in \cite{Cai:2009-PRD,Cai:2009-PLB,EK:2010,Iran,Lu:2014}. In \cite{Chen:2011} the authors studied phase transition of charged topological HL black hole with general dynamical parameter $\lambda$ and phase transition of Kihagias-Sfetosos (KS) black hole. In \cite{Majhi:2012}, phase transition and scaling behavior of charged HL black holes with $\lambda=1$ were studied. Through thermodynamic metrics, in \cite{Mo} the authors analyzed a unified phase transition of the charged topological HL black hole in the case of $\lambda=1$.

In this paper, we will analyze the thermodynamic stability and the phase transition of topological HL black holes in the framework of horizon  thermodynamics.
The previous studies on phase transition of topological HL black hole all referred to concrete matter fields, such as the electromagnetic field, and considered the detailed-balance condition, $\epsilon=0$. We do not care about the concrete matter content, which is exactly one advantage of horizon thermodynamics. Also, we do not employ the explicit form of the HL black hole solution.  Besides, we take general $\epsilon$. It is shown in this paper there are some interesting results which are lack in the usual HL black hole with $\eps =0$.

The paper is arranged as follows. In Section II, we simply review the horizon thermodynamics of HL gravity. In Section III we will
analyze the thermodynamic stability of topological HL black holes. The $P-V$ criticality of HL black hole is discussed in Section IV. We  make some
concluding remarks in Section V.

\section{horizon thermodynamics of topological HL gravity}

The action of HL gravity without the detailed-balance condition is \cite{Lu:2009,Cai:2009-PRD}:

\begin{eqnarray}
\label{action}
I &=& \int dtd^3x ({\cal L}_0 +(1-\epsilon^2){\cal L}_1 +{\cal L}_m),\nonumber \\
 {\cal L}_0 &=& \sqrt{g}N \left \{\frac{2}{\kappa^2}
(K_{ij}K^{ij}-\lambda K^2) +\frac{\kappa^2\mu^2 (\Lambda
R-3\Lambda^2)}{8(1-3\lambda)}\right \},  \nonumber \\
 {\cal L}_1  &=& \sqrt{g}N \left \{\frac{\kappa^2\mu^2(1-4\lambda)}{32(1-3\lambda)}R^2
-\frac{\kappa^2}{2\omega^4}Z_{ij}Z^{ij} \right\},
\end{eqnarray}
where $Z_{ij}=\left(C_{ij}-\frac{\mu
\omega^2}{2}R_{ij}\right)$ with $C_{ij}$ the Cotten tensor. $\epsilon$, $\kappa^2$, $\lambda$, $\mu$, $\omega$ and $\Lambda$ are
constant parameters. ${\cal L}_m$ stands for the Lagrangian of other matter field.

Compared with general relativity, there will be the relations for the parameters:
\be\label{constants}
c=\frac{\kappa^2\mu}{4}\sqrt{\frac{\Lambda}{1-3\lambda}}, \ \
G=\frac{\kappa^2 c}{32\pi}, \ \ \tilde
\Lambda=\frac{3}{2}\Lambda,
\ee
where $G,~c,~\tilde\Lambda$ are Newton's constant, speed of light and the cosmological constant respectively.

Because only for the case with $\lambda=1$, general relativity can be recovered in the large distance approximation. We only consider $\lambda=1$ in the following.
In this case, from Eq.(\ref{constants}), one can see that $\Lambda$ must be negative. Moreover, we will take the natural units: $G=c=1$,
so  $\kappa^2= 32 \pi$ and $\mu^2 = -1/(
 32 \pi^2\Lambda)$.

$\epsilon=0$ is the so-called detailed-balance condition, under which HL gravity turns out to
be intimately related to topological gravity in three dimensions and the geometry of the Cotton tensor.
For the case with $\epsilon=1$, HL gravity returns back to general relativity and HL black hole becomes Schwarzschild-(A)dS black hole. Below we will consider general values of $\epsilon$.

For a static, spherically symmetric black hole, the metric ansatz can be written as
\be\label{staticmetric}
ds^2 =-\tilde N^2(r)f(r) dt^2 +\frac{dr^2}{f(r)} +r^2
d\Omega_k^2,
\ee
where $d\Omega_k^2$ denotes the line element for a two-dimensional
Einstein space with constant scalar curvature $2k$. Without loss of generality, one can take $k=1$ (spherical/elliptic
horizons), $k = 0$ (flat horizons), and $k=-1$ (hyperbolic horizons).

In our work, it is not necessary to know the concrete expression of $f(r)$. Substituting the metric ansatz into the action and varying the action with respect to $\tilde N$,
 the field equation can be obtained \cite{Cai:2010}
 \bea\label{eom}
&& \alpha \left [ 3x^2_++2k-(1-\epsilon^2)
\frac{k^2}{x^2_+}-2x_+ f'-(1-\epsilon^2)\frac{2kf'}{x_+} \right ] \nonumber \\
&&= \frac{16\pi \alpha x^2_+}{\Lambda} \left.\frac{\delta
(\tilde N {\cal L}_m)}{\delta \tilde N}\right|_{x=x_+},
\eea
where $x_{+}=\sqrt{-\Lambda}r_{+}$ and $r_{+}$ is the horizon radius. The $(')$ in $f'$ represents the derivative with respect to $x_{+}$. Here we take a new parameter $\alpha= \frac{\kappa^2\mu^2\sqrt{-\Lambda}\Omega_k}{16}$. In fact, it is shown that when $\lambda=1$ the function $\tilde N(x_{+})$ is a constant \cite{Cai:2009-PLB}. So we will take $\tilde N=1$ below.

In \cite{Cai:2010}, from the field equation above the authors derived the thermodynamic identity $dE=TdS-PdV$, with

\bea
E&=&\frac{\alpha  \left[k^2 \left(1-\epsilon ^2\right)+2 k x_+^2 +x_+^4\right]}{x_+}, \no \\
S&=&\frac{4 \pi  \alpha \left[2 k \left(1-\epsilon ^2\right) \log x_+ +x_+^2\right]}{\sqrt{-\Lambda }}+S_{0},\no \\
P&=& \left.\delta (\tilde N {\cal L}_m)/\delta \tilde N \right|_{x=x_+}, \label{TQ} \\
V&=& -\frac{16 \pi  \alpha  x_+^3}{3 \Lambda },\no \\
T&=& f'(x_{+})\sqrt{-\Lambda}/4\no \pi.
\eea

From the above results, one can see that matter fields do not affect the explicit forms of internal energy $E$ and the entropy $S$ and $V$. For different matter fields, only the $P$ and $T$  have different explicit forms. However, in the framework of horizon thermodynamics, we do not need the exact expressions of $P$ and $T$.

One thing, which should be noted, is that the integration constant $S_0$ in the entropy
cannot be fixed by some physical considerations. To determine $S_0$, one has to invoke the quantum theory of gravity as argued in \cite{Cai:2009-PRD,Cai:2009-PLB}. For $k=0$ case, the area law of the black hole entropy is recovered if setting $S_0=0$. Thus, for simplicity, we always set $S_0=0$ below (including the cases of $k=\pm 1$).

It also should be noted that the pressure $P$ defined by Eq.(\ref{TQ}) is different from the pressure in the usual thermodynamic systems. Especially, due to the metric ansatz Eq.(\ref{staticmetric}) and energy conditions, the pressure defined in this way may have opposite sign to the pressure in the usual sense \cite{Ma2}. For example, the electrodynamic field is a kind of positive pressure matter. However, by Eq.(\ref{TQ}), $P=\frac{\text{$\Lambda $Q}^2}{32 \pi  \alpha ^2 x_+^4}<0$. There is also exotic matter, such as dark energy which supports our universe accelerating expansion, which may have negative pressure in the usual sense. In horizon thermodynamics, by Eq.(\ref{TQ}) they may be positive. Because we require negative $\Lambda$ in this paper, we always consider black holes in the asymptotically AdS spacetime. 
Our discussion does not refer to concrete matter content, in other words, all kinds of matter can be considered as the source. Therefore, there always exist certain matter fields which can give the asymptotically AdS spacetime under appropriate parameter. For example, for the electromagnetic field any negative pressure can guarantee the negative cosmological constant and thus the asymptotically AdS spacetime.  And as mentioned above, we guess that there is also certain exotic matter which corresponds to the positive pressure. Whatever matter sources, the pressure $P$ is negative or positive. Below, under reasonable thermodynamic requirements, such as positive entropy and positive temperature, we will consider the cases with negative and positive pressure.

\section{Thermodynamic stability of topological HL black holes}

One can see that Eq.(\ref{eom}) is in fact an equation of state described by three state parameters $P,~V,~T$, which is
\bea\label{PT}
P&=&\frac{1}{{16 \pi  x_+^4}} \left[ { k^2 \Lambda  \left(\epsilon ^2-1\right)-8 \pi  k \sqrt{-\Lambda } T x_+ \left(\epsilon ^2-1\right) }\right.\no \\
  &+& \left.{ 2 k \Lambda  x_+^2+8 \pi  \sqrt{-\Lambda } T x_+^3+3 \Lambda  x_+^4  }\right]. %
\eea
As noted above, pressures can be positive or negative, while the entropy and the temperatures of black holes must be positive
\footnote{Although there are some physically acceptable systems which have negative temperatures, for black holes the same
explanation does not work.}.
From Eq.(\ref{PT}), we can derive the expression for temperature:
\bea\label{TP}
T=-\frac{k^2 \Lambda  \left(\epsilon ^2-1\right)+2 k \Lambda  x_+^2+x_+^4 (3 \Lambda -16 \pi  P)}{8 \pi  \sqrt{-\Lambda } x_+ \left(-k \epsilon ^2+k+x_+^2\right)}.
\eea
If the Maxwell field exists, the pressure $P$ is given by
\be
P=\frac{\text{$\Lambda $Q}^2}{32 \pi  \alpha ^2 x_+^4}<0.
\ee
The temperature above becomes
\be
T=\frac{\sqrt{-\Lambda } \left(2 \alpha ^2 k^2 \left(\epsilon ^2-1\right)+4 \alpha ^2 k x_+^2-Q^2+6 \alpha ^2 x_+^4\right)}{16 \pi  \alpha ^2 x_+ \left(-k \epsilon ^2+k+x_+^2\right)},
\ee
which is just the result given in \cite{Cai:2009-PRD}.

Positive temperature and positive entropy are only necessary conditions for HL black hole to be in thermodynamic equilibrium. To be in stable equilibrium, $C_P\geq C_V\geq 0$ must be satisfied \cite{Callen}.
According to general definition, for the topological HL black holes
\be
C_{V}=\left.\d{\p E}{\p T}\right|_{V}=\left.T\d{\p S}{\p T}\right|_{V}=0,
\ee
because constant $V$ means constant $E$ and $S$.
We can only define the heat capacity at constant pressure, which is
\bea
C_{P}&=&\left.\d{\p H}{\p T}\right|_{P}=\left.T\d{\p S}{\p T}\right|_{P}\no \\
     &=&\frac{ 8 \pi  \alpha  \left(-k \epsilon ^2+k+x_+^2\right)^2}{A(x_+)} \left[ { k^2 \Lambda  \left(\epsilon ^2-1\right) }\right. \no \\
     &+& \left.{ 2 k \Lambda  x_+^2+x_+^4 (3 \Lambda -16 \pi  P)}\right],
\eea
where $H=E+PV$ is the enthalpy of the system and
\bea
A(x_{+})&=&\sqrt{-\Lambda } \left\{ {k^3 \Lambda  \left(\epsilon ^2-1\right)^2-5 k^2 \Lambda  x_+^2 \left(\epsilon ^2-1\right)}\right. \no \\
       &+& \left.{ k x_+^4 \left[48 \pi  P \left(\epsilon ^2-1\right)+\Lambda  \left(7-9 \epsilon ^2\right)\right]} \right. \no \\
       &+& \left.{x_+^6 (3 \Lambda -16 \pi  P)}\right\}.
\eea
Moreover, according to the viewpoint of Davies \cite{Davies}, the divergence of heat capacity means a phase transition happens there.
It is shown in \cite{Majhi:2012}, when $\epsilon=0$ there will be no phase transition for the charged topological HL black holes with $k=0$ and $k=1$. While there is a second-order phase transition with $k=-1$.

To discuss the global stability of the HL black hole, we also need to calculate the Gibbs free energy, which is defined as
\bea
G&=&E-TS+PV \no \\
 &=&\frac{\alpha  \left[k^2 \left(1-\epsilon ^2\right)+2 k x_+^2+x_+^4\right]}{x_+}-\frac{16 \pi  \alpha  P x_+^3}{3 \Lambda }\no \\
 &-&\frac{\alpha  \left[2 k \left(1-\epsilon ^2\right) \log \left(x_+\right)+x_+^2\right]}{2 \Lambda  x_+ \left(-k \epsilon ^2+k+x_+^2\right)}B(x_{+}),
\eea
where
\be
B(x_{+})=k^2 \Lambda  \left(\epsilon ^2-1\right)+2 k \Lambda  x_+^2+x_+^4 (3 \Lambda -16 \pi  P). \no
\ee
Generally, states with smaller $G$ are more thermodynamically stable.

Below we will first study the entropy and the temperatures of topological HL black holes with general matter fields in cases of $k=0,1,-1$ respectively and with general $\epsilon$.
Then the corresponding heat capacity and Gibbs free energy are calculated and depicted to analyze thermodynamic stability. In the following we take $\Lambda=-1$ and $\alpha=1$ for simplicity.

\subsection{The flat case: $k=0$}

\medskip

In this case, the temperature is proportional to the horizon radius:
\be
T=\frac{(16 \pi  P+3) x_+}{8 \pi }.
\ee
Obviously, if only  $P>-3/16\pi$, the temperature is positive.

The entropy in this case is
\be
S=4 \pi  x_+^2,
\ee
which is always positive.

The heat capacity at constant pressure is
\be
C_P=8 \pi  x_+^2,
\ee
which is also always positive. In addition, its value is independent of $P$, namely the matter fields. So when $k=0$, the HL black hole is always thermodynamically stable and
no phase transition happens.

The Gibbs free energy is
\be
G=-\frac{ 16 \pi  P+3}{6}  x_+^3.
\ee
Because the temperature has to be positive, the Gibbs free energy is always negative and monotonically decreases as the horizon radius grows.

\bigskip

\subsection{The spherical case: $k=1$}

\medskip

In this case, the entropy becomes
\be\label{Sk1}
S=4 \pi  \left[2 \left(1-\epsilon ^2\right) \log x_++x_+^2\right].
\ee
Due to the $\epsilon^2$ in this expression, we only need to analyze the positive $\epsilon$ case. The results with negative $\epsilon$ are the same. Black holes, as thermodynamic systems, should have positive entropy. Thus
the cases with negative entropy will be excluded from thermodynamic consideration.

\begin{figure}
\center{ \subfigure[]{ \label{S1-a}
\includegraphics[width=7cm,keepaspectratio]{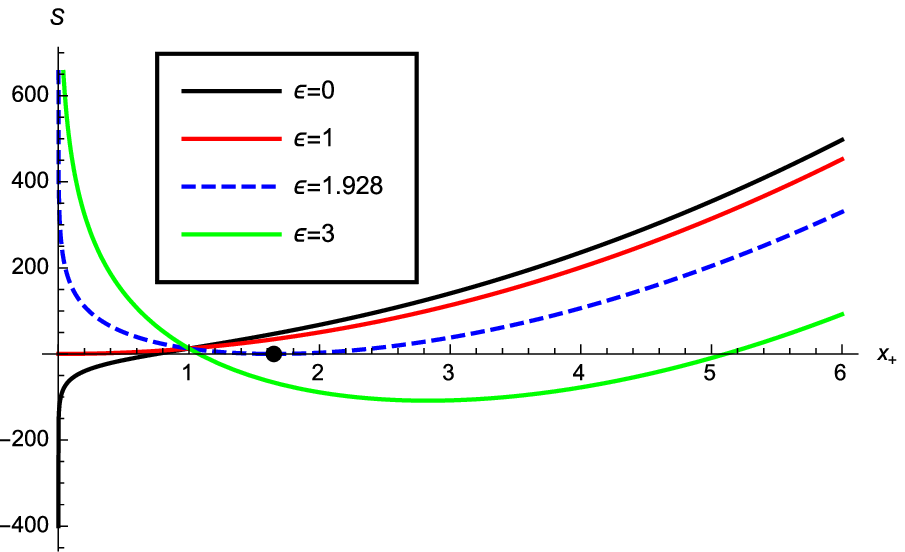}}\\
\subfigure[]{ \label{S1-b}
\includegraphics[width=7cm,keepaspectratio]{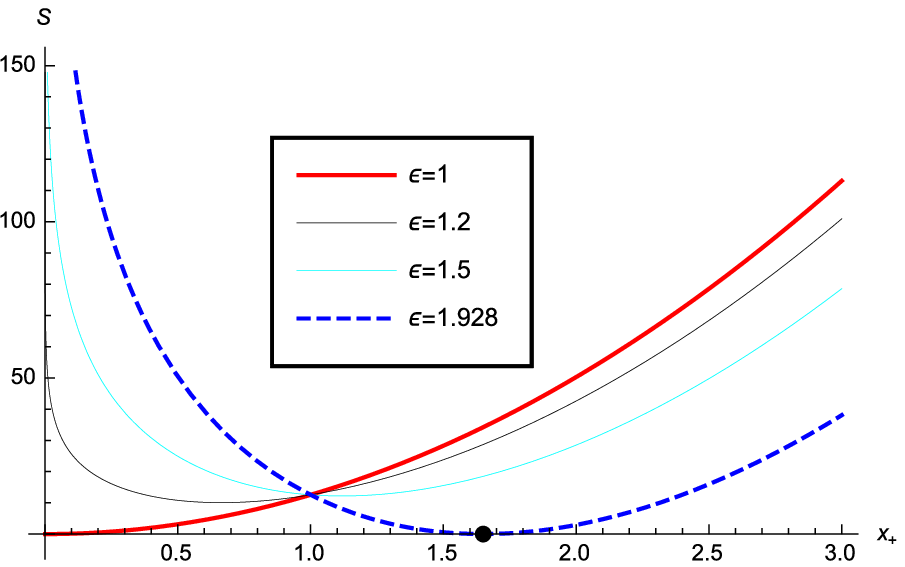}}
\caption{The entropy of topological HL black hole with respect to $x_{+}$ for $k=1$. (b) is the magnification of (a). The black dot lies at $x_{+}=1.649$. At the point $(1,4\pi)$, all the curves cross.   }\label{Sxk1}}
\end{figure}

From Eq.(\ref{Sk1}), we find that there are three cases according to the values of $\epsilon$.
When $0\leq \epsilon < 1$, the black hole entropy is not always positive in the whole range. Only when $x_{+} \geq e^{-\frac{1}{2}W(\frac{1}{1-\epsilon^2})}$, the entropy is positive\footnote{Here $W(x)$ is the Lambert function defined by general formula: $W(x)e^{W(x)}=x$.}.
When $ 1 \leq \epsilon < 1.928$, the black hole entropy is always positive for $x_{+}>0$. And when $\epsilon \geq 1.928$, the black hole entropy takes positive values except for an intermediate region where the entropy is negative. These behaviors have been displayed in Fig.\ref{Sxk1}. Due to the term $\log x_{+}$ in the black hole entropy, all curves cross the point $(1, 4\pi)$. We are not interested in the black hole with positive entropy in two disconnected regions. So we only concern with the HL black hole with $0 \leq \epsilon \leq 1.928$.

When $k=1$, the temperature is
\be\label{Tkp1}
T=\frac{(16 \pi  P+3) x_+^4+2 x_+^2+\epsilon ^2-1}{8 \pi  x_+ \left(x_+^2-\epsilon ^2+1\right)}.
\ee
Its value depends on the relations between $P$ and $\epsilon$. We find that there always exist some regions where $T>0$ only if $P > \frac{4-3 \epsilon ^2}{16 \pi  \epsilon ^2-16 \pi }$.


\begin{figure}
\center{ \subfigure[$P=-1/6\pi$]{ \label{2-a}
\includegraphics[width=7cm,keepaspectratio]{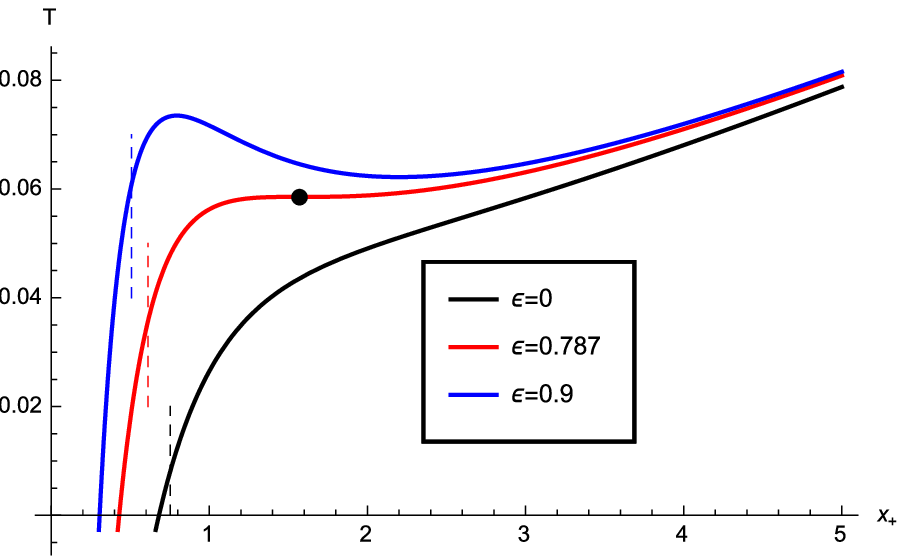}}\\
\subfigure[$P=-1/6\pi$]{ \label{2-b}
\includegraphics[width=7cm,keepaspectratio]{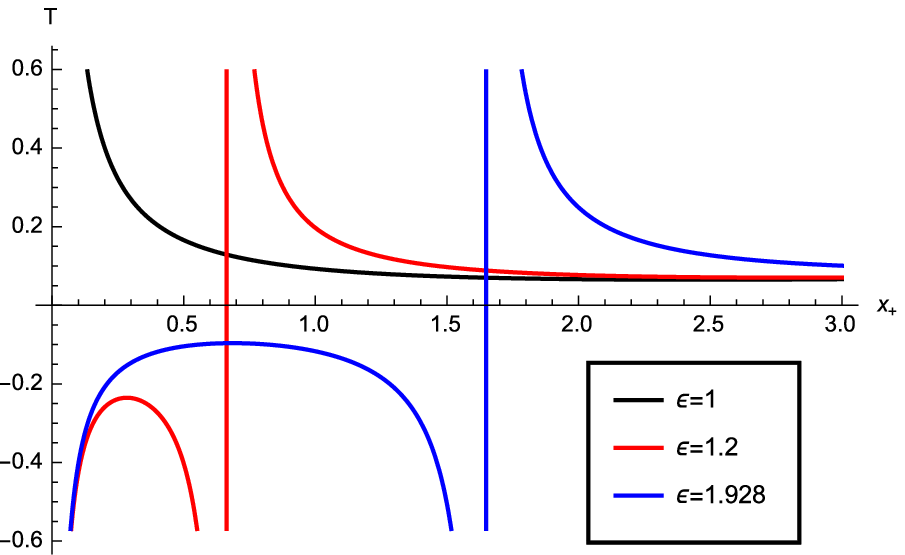}}\\
\subfigure[$P=1/\pi$]{ \label{2-c}
\includegraphics[width=7cm,keepaspectratio]{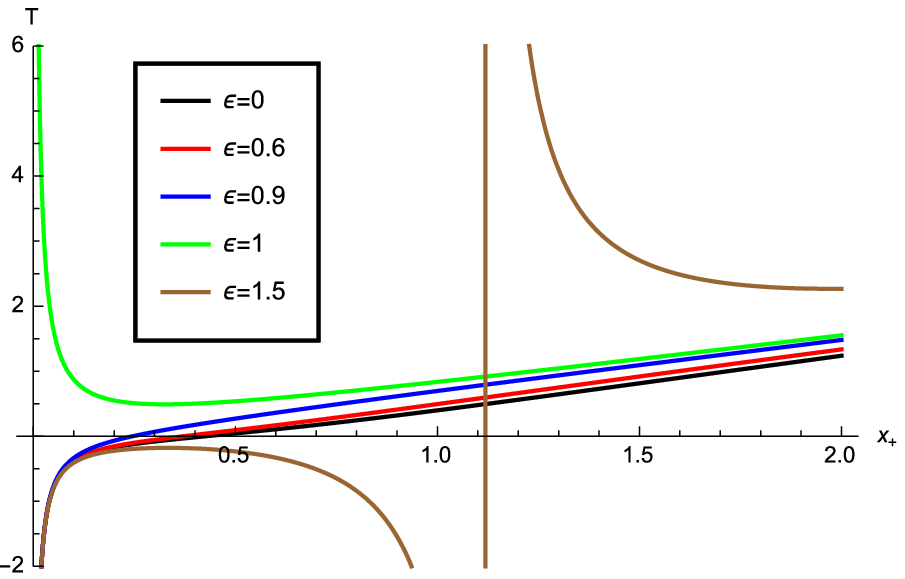}}
\caption{Temperatures of topological HL black hole with respect to $x_{+}$ for $k=1$. In (a), the three vertical dashed lines lie at $x_{+}=0.508,~x_{+}=0.612, ~x_{+}=0.753$ respectively. Only on the right of these lines are the corresponding entropies positive. The black point represents the critical point with $\epsilon=0.787$ and $x_{+}=1.57$. In (b), Only on the right of the divergent points are both the temperature and entropy positive. In (c), the positive pressure case. }\label{Tk1}}
\end{figure}

We plot the temperatures of HL black hole in Fig.\ref{Tk1}.
Interestingly, as shown in Fig.2(a), for negative pressure and $0<\epsilon<1$, the temperature of HL black hole exhibits the similar behavior to that of Reissner- Nordstr\"{o}m (RN)-AdS black hole. Here the parameter $\epsilon$ plays the similar role to the electric charge $Q$ in RN black hole. However, considering the requirement of positive entropy, only in the regions on the right of the vertical dashed lines is the black hole physical. In Fig.2(b), we depict the temperature of HL black hole for $1\leq \epsilon \leq 1.928$. In this case, there are always divergent points due to this term $(x_+^2-\epsilon ^2+1)$ in the denominator of $T$ in Eq. (\ref{Tkp1}). Therefore, only in the regions on the right of the divergent points are the HL black holes with positive temperature physically acceptable. For comparison, we also depict the curves of $T-x_{+}$ in the positive pressure case (Fig. 2(c)).
In this case, for $1\leq \epsilon \leq 1.928$ the temperature increases monotonically as the horizon radius $x_{+}$ grows. Thus, the HL black hole, not like its negative pressure counterpart, does not exhibit the critical behavior. While for $\epsilon>1$, the HL black hole with positive pressure has the similar behavior to the one with negative pressure case. It should be noted that the pressure $P$ with the different signs may correspond to different matter source. For example, the negative pressure cases in Fig.2(a), Fig.2(b) can describe the HL black hole with the electromagnetic field and the positive pressure case in Fig.2(c) describes the temperatures of HL black holes with other matter sources.

\begin{figure}
\includegraphics[scale=0.3]{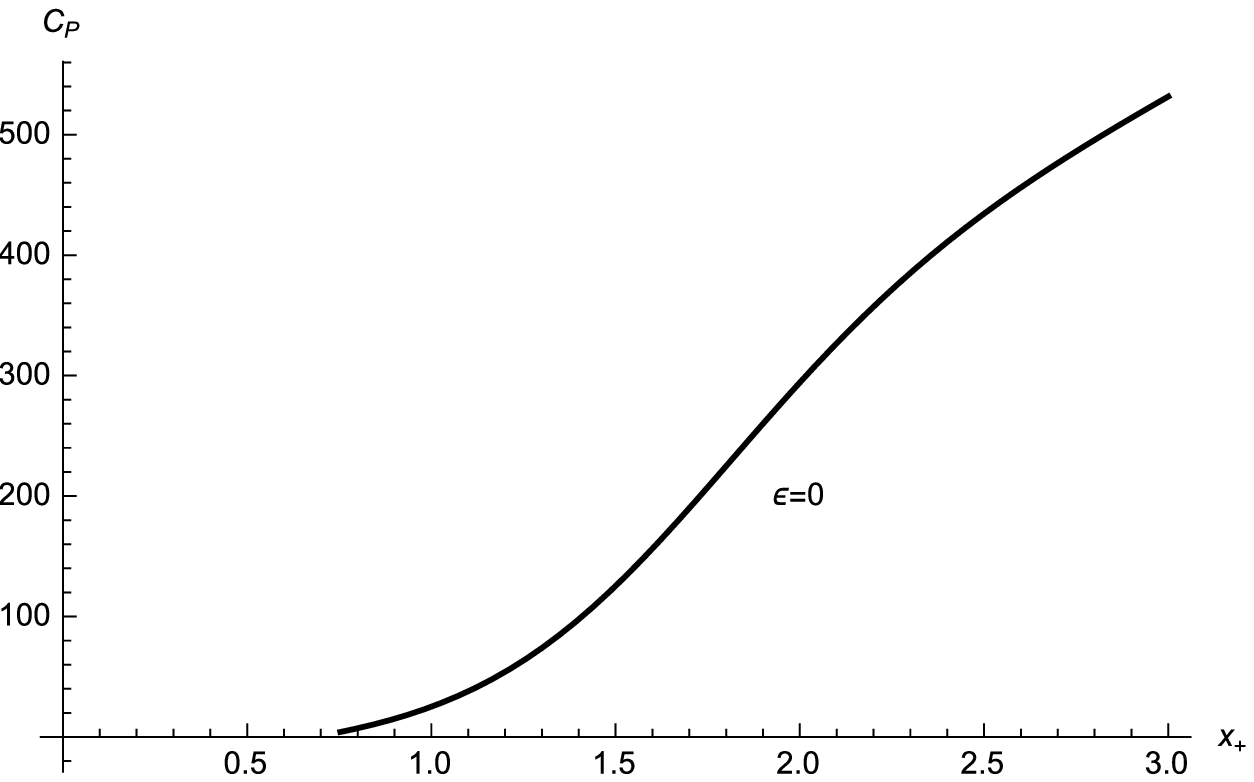}\hspace{0.2cm}
\includegraphics[scale=0.3]{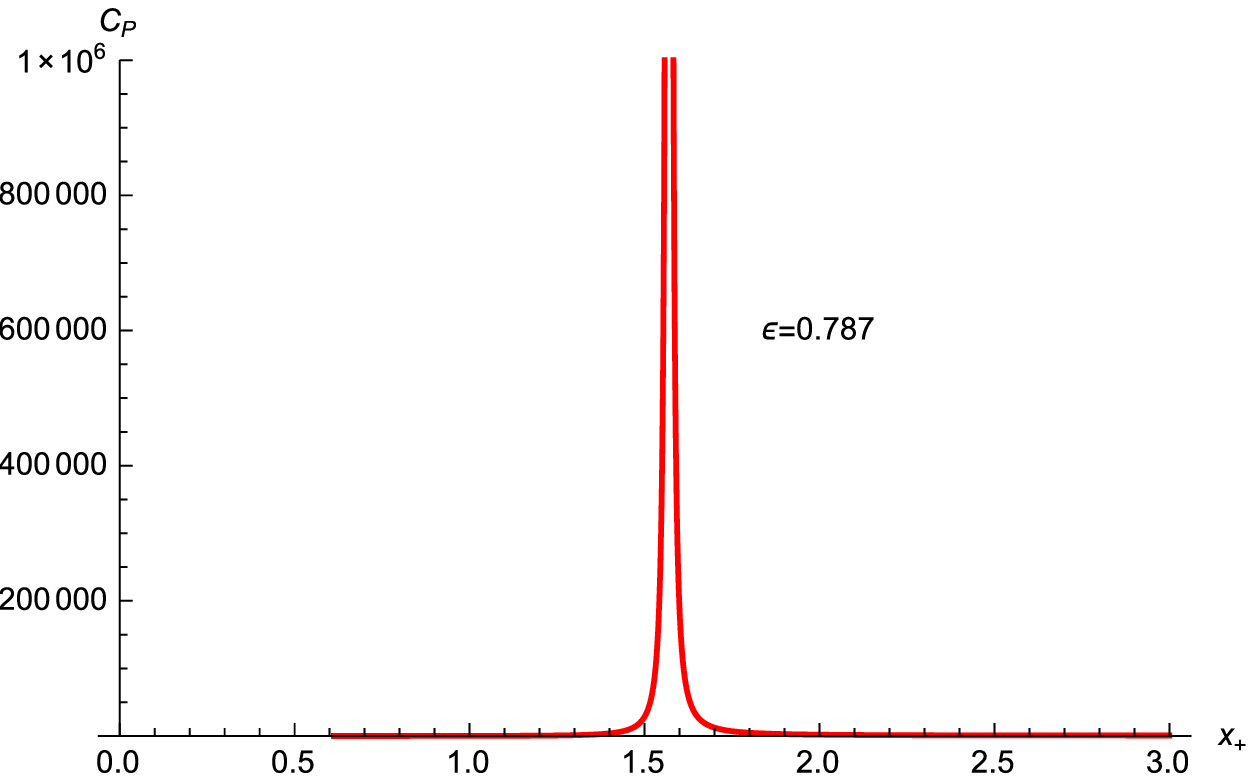}\\
\includegraphics[scale=0.3]{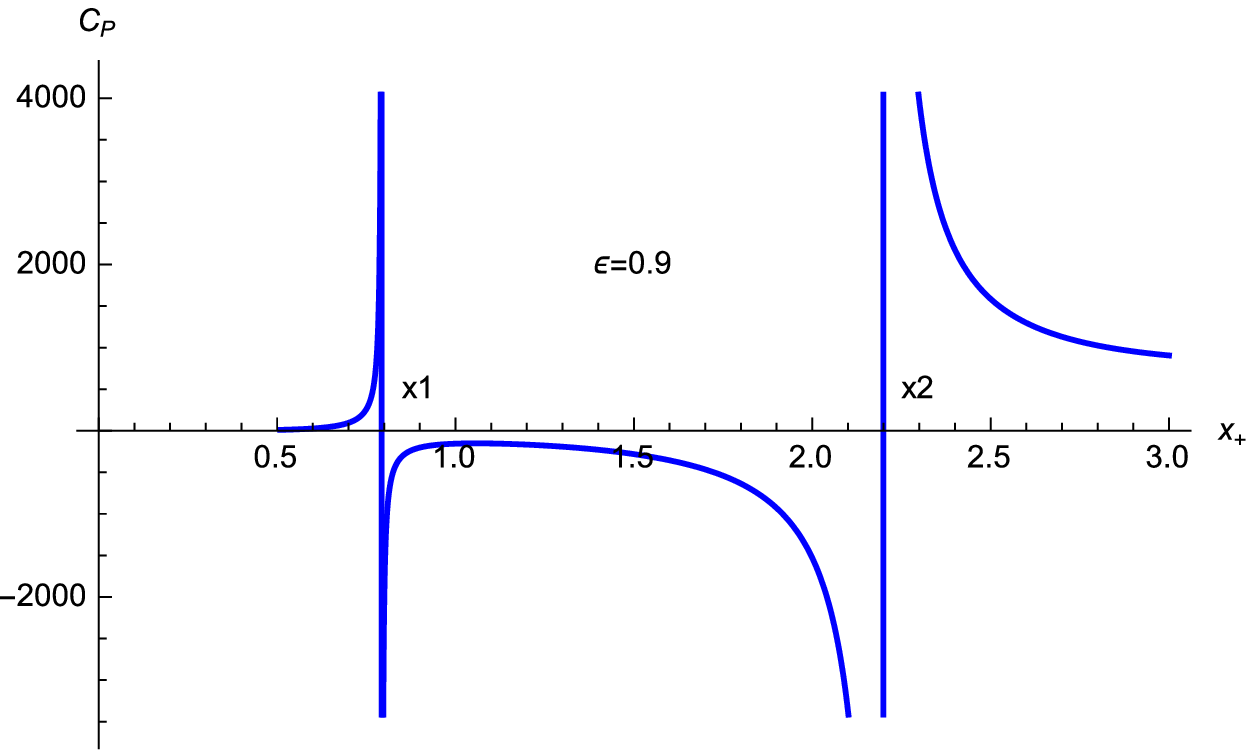}\hspace{0.2cm}
\includegraphics[scale=0.3]{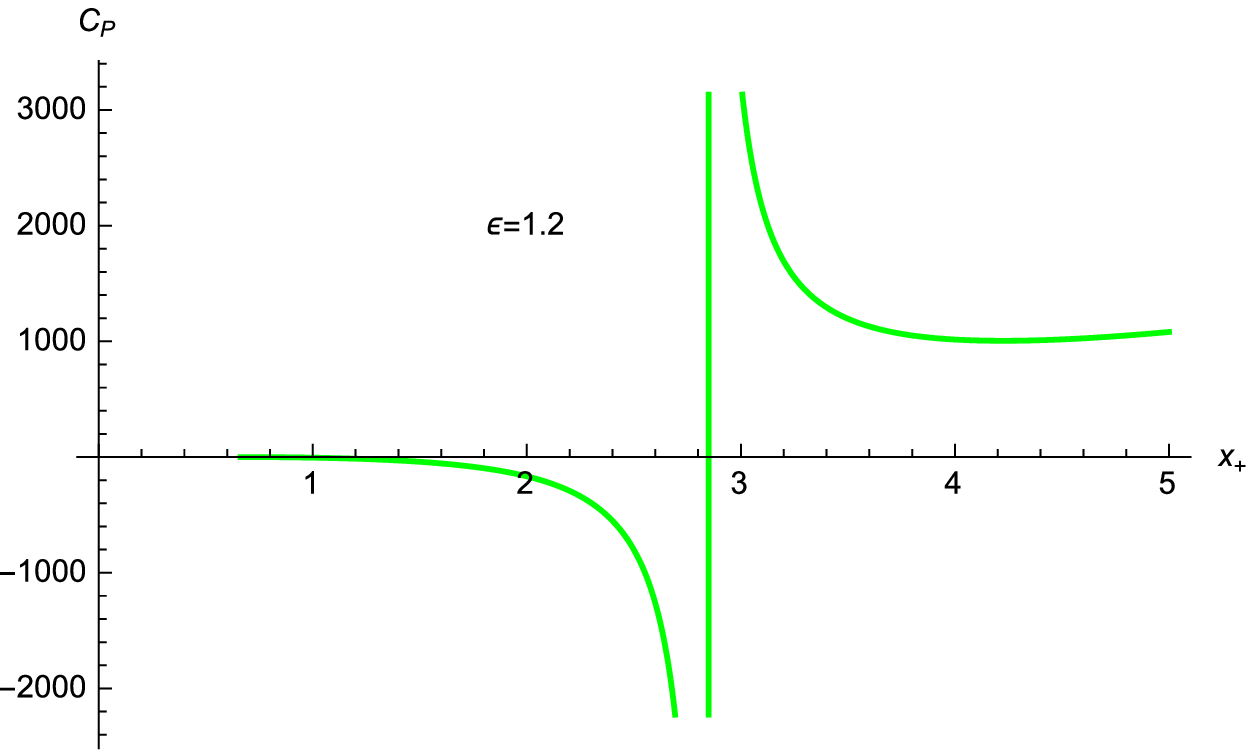}
\caption{Heat capacity at constant pressure $C_P$ of topological HL black hole for $k=1$ with respect to $x_{+}$. The two divergent points in the bottom left diagram are $x1=0.793,~x2=2.199$. We take $P=-1/6\pi$. }\label{CP1}
\end{figure}

Excluding the regions with negative entropy and negative temperature, we depict the heat capacities at constant pressure in Fig.\ref{CP1}.
Clearly, when $0 \leq \epsilon < 0.787 $ the heat capacity is always finite and positive. Due to the positivity of $C_P$, the HL black hole for negative pressure with $0 \leq \epsilon < 0.787 $ is locally thermodynamically stable. When $0.787 < \eps <1$, the heat capacity will suffer discontinuities at two different points $x1,~x2$, which may be identified as the critical points for the phase transition phenomena in HL black holes. The heat capacity $C_P$ is positive for $x_{+}<x1$ and $x_{+}>x2$, while it is negative in the intermediate region $x1<x_{+}<x2$. Therefore, across the two critical points, there is a change of thermodynamic stability of the HL black hole. When the parameter takes the critical value $\eps_{c} = 0.787$, the two critical points $x1,~x2$ coincide. In this case, the intermediate unstable phase disappears and the HL black hole is always thermodynamically stable. For $\eps \geq 1$, there is only one divergent point for $C_P$. In this case, the HL black hole tends to cross the divergent point from the left to the right.
We do not depict the heat capacity for positive pressure because its behaviors are simple. For positive pressure, the $C_P$ is always positive  When $\eps <1$ and when $\eps \geq 1$ there will be a divergent point.

\begin{figure}
\includegraphics[width=7cm,keepaspectratio]{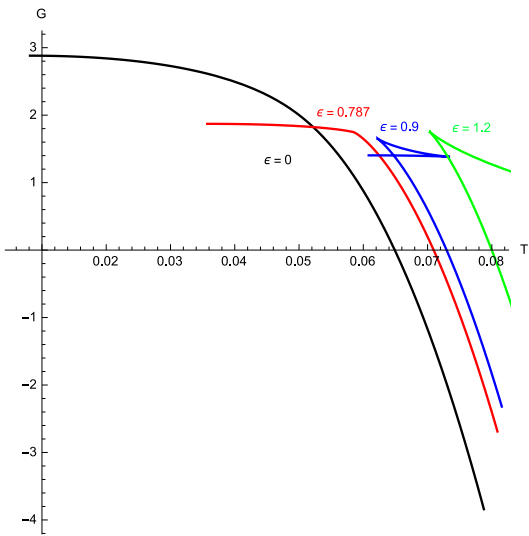}
\caption{Gibbs free energy of topological HL black hole for $k=1$ with respect to $T$. Here we have taken $P=-1/6\pi$.. }\label{G1-a}
\end{figure}

To analyze the global thermodynamic stability of HL black hole, we depict the Gibbs free energy in Fig.\ref{G1-a} as functions of temperature for fixed pressure. As is shown in the figure, the Gibbs free energy develops a  `` swallow tail" behavior when $1>\eps>\eps_c=0.787$, which is the characteristic of first-order phase transition. When $\eps=\eps_c$, the " swallow tail" disappears, corresponding to the critical point. For $\eps >1$, a cusp appears instead of the `` swallow tail". Thus, in this case, the divergent point for $C_P$ is not a real critical point and no phase transition takes place.

\bigskip

\subsection{The hyperbolic case: $k=-1$}

\medskip

\begin{figure}
\includegraphics[scale=0.8]{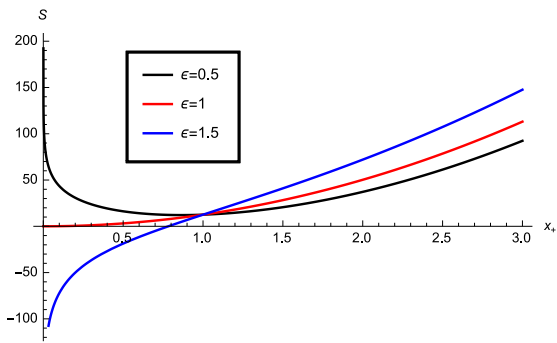}
\caption{The entropy of topological HL black hole with respect to $x_{+}$ for $k=-1$. All the curves also cross the point $(1,4\pi)$.  }\label{Sxkn1}
\end{figure}

Now, the entropy of HL black hole is
\be
S=4 \pi  \left[x_+^2-2 \left(1-\epsilon ^2\right) \log x_+\right].
\ee
The structure of the entropy in the $k=-1$ case is simpler than that of $k=1$ case. As is shown in Fig. \ref{Sxkn1}, for $0\leq \eps < 1$ the entropy is always positive. While it has one zero point for $ \eps > 1$.
This point lies at $x_{+} = e^{-\frac{1}{2}W(\frac{1}{\epsilon^2-1})}$.
On the right-hand side of this point, the entropy is positive.

\begin{figure}
\center{ \subfigure[ $\epsilon=0.5$]{ \label{Tk2-a}
\includegraphics[scale=0.8]{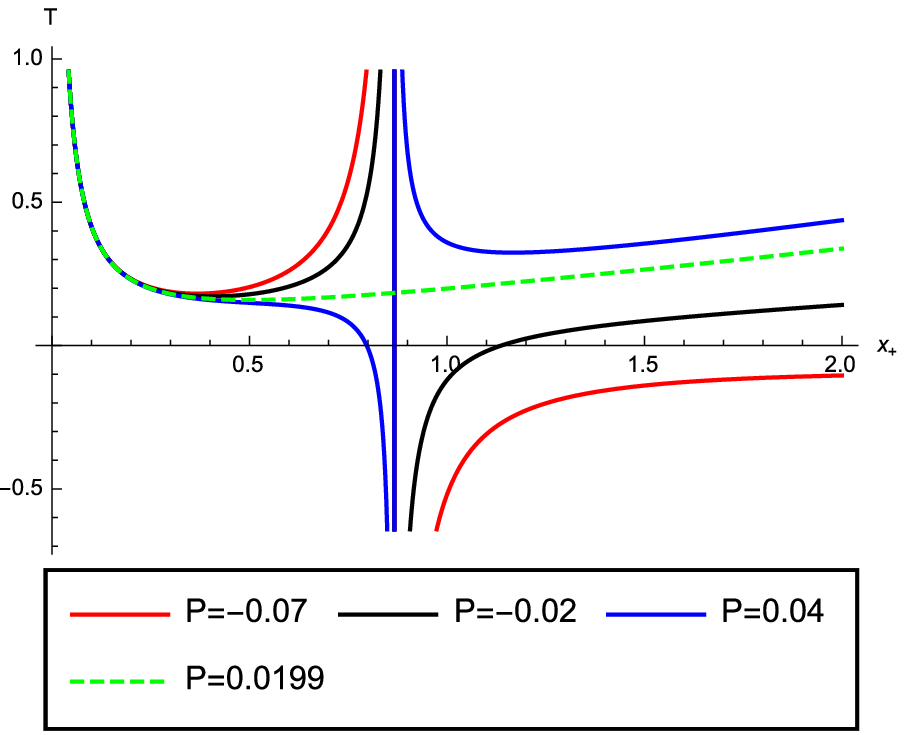}}
\subfigure[ $\epsilon=1.5$]{ \label{Tk2-b}
\includegraphics[scale=0.8]{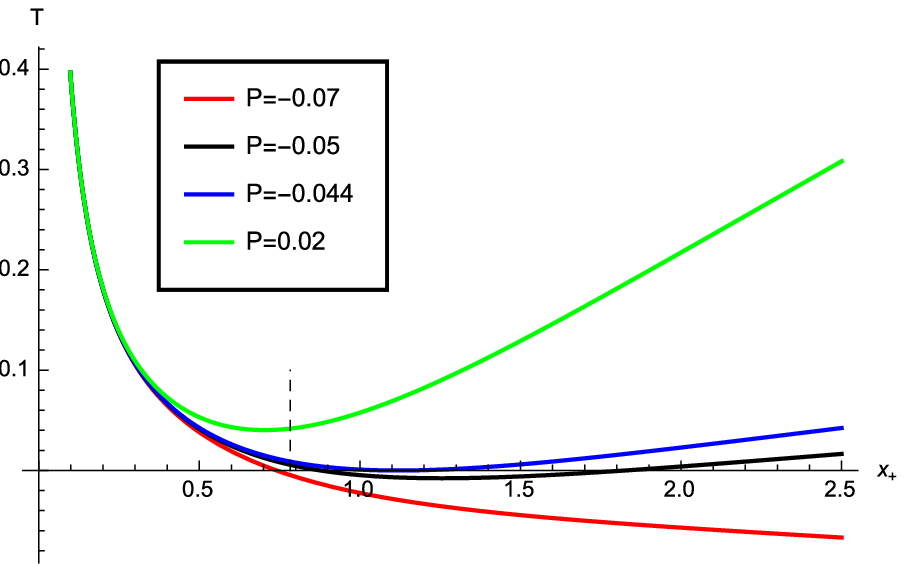}}
\caption{Temperatures of topological HL black hole with respect to $x_{+}$ for $k=-1$. In (a), the dashed (green) curve corresponds to $P=-\frac{3 \epsilon ^2}{16 \pi  \epsilon ^2-16 \pi }$. In (b), the vertical dashed line marks the point on the right-hand side of which the entropy is positive.}\label{Tk2}}
\end{figure}

In this case, the temperature is
\be\label{Tkn1}
T=\frac{(16 \pi  P+3) x_+^4-2 x_+^2+\epsilon ^2-1}{8 \pi  x_+ \left(x_+^2+\epsilon ^2-1\right)}.
\ee
There are six cases in which the temperatures can be positive. They are listed in Table \ref{Tab2}. The corresponding temperatures are depicted in Fig.\ref{Tk2}.
Clearly, when $k=-1$ the behaviors of temperatures are more complicated. In this case, the pressure $P$ has more apparent effects on the temperature $T$. So we will fix the parameter $\eps$ and vary the values of pressure $P$.

\renewcommand\arraystretch{1.5}
\begin{table}[!hbp]
\centering
\begin{tabular}{|c|c|c| }
\hline\hline
$\epsilon$ & $P$ & No. \\
\hline
\multirow{3}*{$|\epsilon| <1$} & $P\leq-\frac{3}{16 \pi }$ & I  \\ \cline{2-3}
                               & $-\frac{3}{16 \pi }<P<-\frac{3 \epsilon ^2}{16 \pi  \epsilon ^2-16 \pi }$ & II  \\ \cline{2-3}
                               & $P\geq-\frac{3 \epsilon ^2}{16 \pi  \epsilon ^2-16 \pi }$ & III \\ \cline{2-3}
\hline
\multirow{3}*{$|\epsilon| >1$} & $P\leq-\frac{3}{16 \pi }$ & IV \\ \cline{2-3}
                               & $-\frac{3}{16 \pi }<P<\frac{4-3 \epsilon ^2}{16 \pi  \epsilon ^2-16 \pi }$ & V \\ \cline{2-3}
                               & $P\geq \frac{4-3 \epsilon ^2}{16 \pi  \epsilon ^2-16 \pi }$ & VI \\ \cline{2-3}
\hline
\end{tabular}
\caption{For $k=-1$, under these conditions the temperature has different behaviors.}\label{Tab2}
\end{table}

According to Eq.(\ref{Tkn1}), one can easily find that there is always another divergent point besides $x_{+}=0$ for $0<\eps<1$ except for a special pressure $P=-\frac{3 \epsilon ^2}{16 \pi  \epsilon ^2-16 \pi }$. When $P$ satisfies the condition (I) in Table \ref{Tab2}, the temperature is positive only on the left-hand side of the divergent point. While for conditions (II, III), the temperature is positive in two disconnected regions. Due to the existence of negative temperature region between them, the black hole cannot transit from one positive temperature region to the other.

When $\eps >1$, under the condition (IV) the temperature monotonically decrease as the horizon radius $x_{+}$ increases. When $P$ satisfies condition (V), the temperature $T$ is positive at two disconnected regions.
When the pressure falls into the scope in condition (VI), the temperature will be always positive. However, considering the positivity of entropy when $\eps >1$, there must be $x_{+} > e^{-\frac{1}{2}W(\frac{1}{\epsilon^2-1})}$.

\begin{figure}
\includegraphics[width=7cm,keepaspectratio]{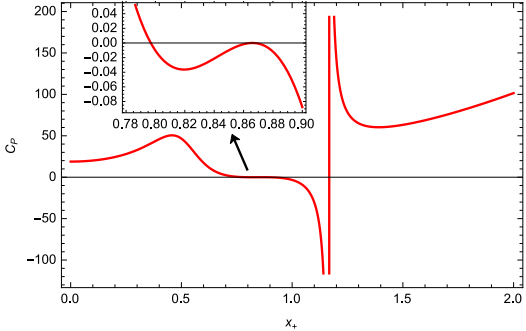}
\caption{Heat capacity at constant pressure $C_P$ of topological HL black hole for $k=-1$ with respect to $x_{+}$. Here $\eps=0.5$ and $P=0.04$. }\label{CP2}
\end{figure}

Among the six cases, we are interested in the case under the condition (III). One can notice from Fig.\ref{Tk2-a} that at most there are three types of black holes, the smaller one, the intermediate one and the larger one, which can have the same thermodynamic state, namely the same $(T,P)$. We want to know which one is the most thermodynamically preferred. We depict the corresponding heat capacity in Fig.\ref{CP2}. There are two zero points for the $C_P$ at the points $x_{+}=0.797$ and $x_{+}=0.866$.  The left one corresponds to the zero point of the temperature and the right one corresponds to the divergent point of the temperature. In the two regions: $0<x_{+}<0.797$ and $x_{+}>0.866$, the heat capacity takes positive values. This means that the smaller black hole and the larger one can be both locally thermodynamically stable. In the region $0.797<x_+<0.866$ the heat capacity $C_P$ is negative. So the HL black hole in this region is not thermodynamically stable and will transit to the smaller black hole in the region $0<x_{+}<0.797$ or to the larger black hole in the region $x_{+}>0.866$ through a phase transition. It is well known that under the same pressure and temperature a thermodynamic system will be in a state with smallest chemical potential (Gibbs free energy). Although the intermediate black hole would be disregarded due to its negative heat capacity, we still need the Gibbs free energy to ascertain the globally thermodynamic stable state for the other two cases.

\begin{figure}
\includegraphics[width=6cm,keepaspectratio]{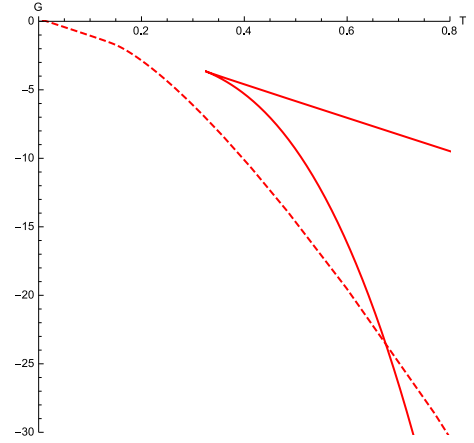}
\caption{Gibbs free energy of topological HL black hole for $k=-1$ with respect to $T$. The dashed curve corresponds to the smaller black hole in the range $x_{+}<0.797$ and the solid curve corresponds to the larger black hole with $x_{+}>0.866$. The cross point of the two curves is $T=0.673,~G=-23.4$. Here $\eps=0.5$ and $P=0.04$.}\label{G2}
\end{figure}

  As is shown in Fig.\ref{G2}, when the temperature is below $T=0.673$, the thermodynamic state of the smaller black hole has the lower Gibbs free energy. While the larger black hole is thermodynamically preferred when the temperature is above $T=0.673$. However, due to the cutoff of the negative temperature region, there will be no phase transition for the HL black hole from one region to the other.

\section{P-V criticality of topological HL black holes}

Almost all the previous works on $P-V$ criticality considered the cosmological constant $\Lambda$ to be variable and treated it as the thermodynamic pressure. In this way, it has been verified that the usual charged HL black hole do not have the $P-V$ criticality \cite{MoJ}. In the present paper, we employ the horizon thermodynamics and treat the energy-momentum tensor of matter fields as the pressure. According to Eq.(\ref{PT}), we can discuss the $P-V$ criticality of HL black holes. Below we still fix $\Lambda=-1$. It shows that in horizon thermodynamics, $P-V$ criticality can exist.

\medskip

\textbf{(1)}. $k=0$
\medskip

In this case, the equation of state is
\be
P=-\frac{3}{16 \pi }+\frac{T}{2 x_+}.
\ee
Clearly, no criticality exists.

\bigskip

\textbf{(2)}. $k=-1$
\medskip

\begin{figure}[!htbp]
\includegraphics[width=7cm,keepaspectratio]{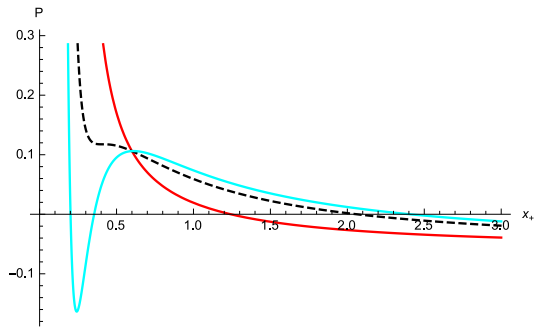}
\caption{The $P-V/x_{+}$ diagram of topological HL black hole for $k=-1$. The temperature of isotherms grows from top to bottom on the left of the point $x_{+}=\sqrt{1-\epsilon^2}$ and decreases from top to bottom on the right of that point. The critical isotherm is depicted by the dashed (black) line. We have taken $\Lambda=-1$ and $\epsilon=0.8$.}\label{PV2}
\end{figure}

In this case, the equation of state is
\bea\label{PT2}
P&=&\frac{1}{{16 \pi  x_+^4}} \left[ { \left(1-\epsilon ^2\right)+8 \pi T x_+ \left(\epsilon ^2-1\right) }\right.\no \\
  &+& \left.{ 2 x_+^2+8 \pi  T x_+^3 - 3 x_+^4  }\right]. %
\eea
By comparing the above equation with the van der Waals equation, one can easily find the specific volume $v \propto x_{+}$.
So we would not introduce the specific volume, but directly study the $P-x_{+}$ behavior.
The behavior of this equation is displayed in Fig.\ref{PV2}. Clearly, in this case the HL black hole has no $P-V$ criticality. However, another interesting thermodynamic behavior exists. There is an interesting point where all isotherms intersect. This indicates that this point is independent of the temperature.
It originates from the counterbalance of the second term and the fourth term in Eq.(\ref{PT2}). When $ (1-\epsilon ^2)=x_+^2$, the temperature is cancelled out in Eq.(\ref{PT2}).
This special point is usually called `` thermodynamic singularity" \cite{FAM}. It is characterized by
\be\label{thermosingu}
\left.\frac{\p P}{\p T}\right|_{V=V_s}=0,
\ee
which precisely gives $ (1-\epsilon ^2)=x_+^2$. And indeed, the temperature and Gibbs free energy both diverge at the thermodynamic singularity apart from a very special point $P= P_s=-\frac{3 \epsilon ^2}{16 \pi  \epsilon ^2-16 \pi }$ for which the temperature and the Gibbs free energy are both finite. This special pressure is just the lower bound of Condition (III) we give in Table.I. The temperature for $P_s$ has also depicted in Fig. \ref{Tk2-a}, which is indeed finite. For Gauss-Bonnet black hole and Lovelock black hole, there are also the similar thermodynamic singularities and have been extensively analyzed in \cite{FAM}.

\bigskip

\textbf{(3)}.$k=1$

\medskip

\begin{figure}[htp]
\includegraphics[width=7cm,keepaspectratio]{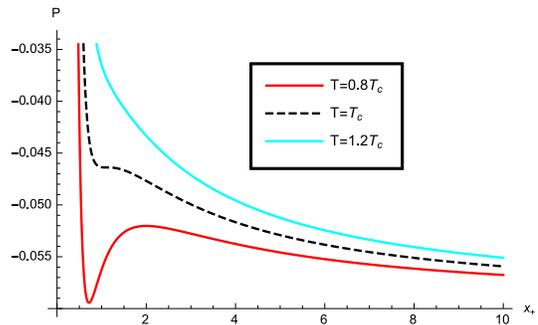}
\caption{The $P-V/x_{+}$ diagram of topological HL black hole for $k=1$. The temperature of isotherms decreases from top to bottom. The critical isotherm is depicted by the dashed (black) line. The critical temperature $T_c=0.829$. We have taken  $\epsilon=0.9$.}\label{PV1}
\end{figure}

In this case, Eq.(\ref{PT}) turns into
\bea\label{PT1}
P&=&\frac{1}{{16 \pi  x_+^4}} \left[ { \left(1-\epsilon ^2\right)-8 \pi T x_+ \left(\epsilon ^2-1\right) }\right.\no \\
  &-& \left.{ 2 x_+^2+8 \pi  T x_+^3 -3 x_+^4  }\right]. %
\eea
The behavior of this equation is displayed in Fig.\ref{PV1}. Apparently, in this case the HL black hole has the similar $P-V$ criticality to those of RN-AdS black hole and van der Waals liquid/gas system.

The critical point lies at:
\bea
T_c&=&\frac{\left(2-\sqrt{3}\right) \sqrt{3+2 \sqrt{3}}}{6 \pi  \sqrt{1-\epsilon ^2}}, \no \\
P_c&=&\frac{-27 \epsilon ^2+8 \sqrt{3}+12}{144 \pi  \left(\epsilon ^2-1\right)}, \no \\
x_{c}&=&\sqrt{3+2 \sqrt{3}} \sqrt{1-\epsilon ^2}.
\eea
Clearly, the $T_c,~x_c$ will be imaginary if $\eps^2>1$. Therfore, the critical behavior only exists when $\epsilon^2<1$. Correspondingly, one can calculate
\be
\rho_c=\frac{P_c ~x_c}{T_c}=\frac{-27 \epsilon ^2+8 \sqrt{3}+12}{24 \left(\sqrt{3}-2\right)}.
\ee
It only depends on the parameter $\epsilon$ and has nothing to do with $\Lambda$ (Even if we do not take a special value for $\Lambda$, it can be eliminated from this relation.).  Generally $\rho_c$ is negative. Only when $\epsilon^2>0.96$, $\rho_c$ can be positive. So it is very different from the
usual van der Waals liquid/gas system.

\begin{figure}[!htbp]
\includegraphics[width=7cm,keepaspectratio]{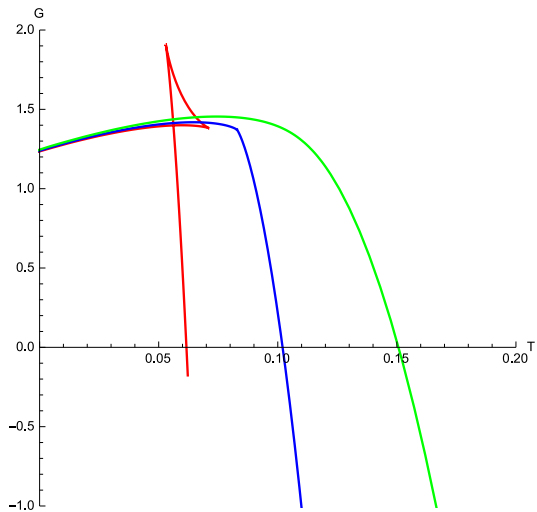}
\caption{The $G-T$ diagrams of topological HL black hole for $k=1$ for fixed $\eps=0.9$. The blue curve corresponds to the critical pressure $P=P_c$, and the red curve corresponds to the pressure under the $P_c$. }\label{GTP1}
\end{figure}

Defing
\be
t=\d{T}{T_c}-1, \quad \zeta=\d{x_{+}}{x_c}-1, \quad p=\d{P}{P_c}
\ee
and replacing $x_{+},~T,~P$ in Eq.(\ref{PT1}) with the new dimensionless parameters $\zeta,~t,~p$ and then expanding the equation near the critical point approximately, one can obtain
\be\label{app}
p=1+At+Btx+Cx^3+O(tx^2,x^4),
\ee
where $A,~B,~C$ are all functions of $\eps$ . Eq.(\ref{app}) has the same form as that for the van der Waals system and the RN-AdS black hole \cite{Mann1}. Therefore, for this system
the critical exponents should also be $\beta=1/2,~\gamma=1,~ \delta=3$. In addition, because $C_V=0$, we also have the critical exponent $\alpha=0$.
Obviously, they obey the scaling symmetry like the ordinary thermodynamic systems.

As is depicted in Fig. \ref{GTP1}, when we fix the value of $\eps$, the $G-T$ diagrams also exhibit `` swallow tail" behavior for $P<P_c$, which means a first-order phase transition exists.

In horizon thermodynamics, the phase structure of black hole is comparatively simple.  We find that the HL black hole does not have other interesting critical behaviors, like isolated critical point \cite{Dolan2014} or triple point \cite{NA,LYX2}.

In fact, for the spherical HL black hole there is still the thermodynamic singularity. Only if $\epsilon^2>1$, the second term and the fourth term in Eq.(\ref{PT1}) can also offset and the residual equation is independent the temperature $T$. However, the cost for the thermodynamic singularity is the disappearance of the $P-V$ criticality.

%

\section{Concluding remarks}

In the framework of horizon thermodynamics, we studied the thermodynamic stabilities of the topological black holes in HL gravity. In static, spherically symmetric case, whatever the matter fields are, one can always set the $(rr)$ component of energy-momentum tensor as the  pressure of black hole thermodynamic system to obtain the thermodynamic identity $dE=TdS-PdV$ from the field equations. We study the thermodynamic stabilities of black holes according to the variables $E,~T,~S,~P,~V$ by taking the same method and criterion as those in the usual thermodynamics. The concrete matter fields are not important. Also, the concrete metric function is not necessary to know. Therefore, our discussion applies for all the black holes in the gravitational theories under consideration.

For topological HL black holes, the entropy and the temperatures are not always positive. Negative entropy and negative temperature are meaningless for black holes. We should analyze the thermodynamic stability of topological HL black holes only in the regions with positive entropy and temperature. It is found that the temperature of topological HL black holes for $k=0$ can be positive when $P>-3/16\pi$. And the heat capacity at constant pressure is always positive, which means the black hole in this case is always thermodynamically stable and does not have any phase transition. For $k=1$, we fixed the pressure and studied the influence of the parameter $\eps$ on the thermodynamic behaviors of the HL black hole. We find that for some $\eps$ the HL black hole exhibits the similar thermodynamic behavior to that of RN-AdS black hole. Its temperature has two extrema. Correspondingly, the $G-T$ diagrams of HL black hole exhibit the `` swallow tail" behavior which characterize the first-order phase transition. For $k=-1$, the behavior of entropy is simple, however the temperature has fruitful structures. There are six cases in which there are positive temperatures. In this case, the pressure has more apparent effects on the thermodynamic behaviors of HL black hole. So we fixed the $\eps$ and varied the values of pressure $P$. We mainly focus on the case under condition (III) in Table.I. In this case, the black hole can have at most three configurations with the same thermodynamic states. We analyze which one of the three is thermodynamically preferred according to heat capacity and Gibbs free energy. However, due to the existence of negative temperature region between them, the black hole cannot transit
from one positive temperature region to the other.

We also studied the $P-V$ criticality of topological HL black hole in the framework of horizon thermodynamics through the equation of state $P=P(V,T)$. We find that the $P-V$ diagram of topological HL black hole for $k=1$ looks the same as that of van der Waals liquid/gas system. They also have the same critical exponents. However, the parameter $\rho_c=\frac{P_c ~r_c}{T_c}$ of the two systems is very different. When $k=-1$, no $P-V$ criticality exists. But there is a special point called `` thermodynamic singularity"  in the  $P-V$ diagram, where the isotherms intersect. On the left and right side of the point, the isotherms have opposite sequence. This singularity is characterized by $\left.\frac{\p P}{\p T}\right|_{V=V_s}=0$. At the `` thermodynamic singularity",  the temperature and Gibbs free energy are both divergent apart from a special pressure $P=P_s$ where the temperature and the Gibbs free energy are finite.

It may reflect some universality of the thermodynamic behaviors of HL black holes in the framework of horizon thermodynamics. However, it may miss some peculiarity of the topological HL black holes. It would be plausible to further study the thermodynamic stability and  critical behaviors of the HL black hole with some concrete matter field like the electromagnetic field, which may further shed light on the thermodynamic properties of HL black holes.

\bigskip

\section*{Acknowledgements}
This work is supported in part by the National Natural Science Foundation of China under Grants
 Nos.(11605107, 11475108) and by the Doctoral Sustentation Fund of Shanxi Datong
University (2011-B-03).


\begin{thebibliography}{99}

\bibitem{HP}S. Hawking, D.N. Page, Commun. Math. Phys. 87, 577(1983).
\bibitem{Chamblin}A. Chamblin, R. Emparan, C. Johnson, and R. Myers, \PRD 60 ,064018(1999); Phys. Rev. D 60, 104026 (1999).
\bibitem{Lemos}C. S.Peca, J. P. S. Lemos, \PRD 59, 124007(1999).
\bibitem{Wu}X. N. Wu, \PRD 62, 124023(2000).
\bibitem{MMC}M. M. Caldarelli, G. Cognola, and D. Klemm, \CQG 17, 399 (2000).
\bibitem{Kastor}D. Kastor, S. Ray, and J. Traschen, \CQG 26, 195011 (2009).
\bibitem{Dolan}B. P. Dolan, \CQG 28, 125020 (2011).
\bibitem{Mann1}D. Kubiznak and R. B. Mann, J. High Energy Phys. 07(2012) 033.
\bibitem{Mann2}S. Gunasekaran, D. Kubiznak, and R. B. Mann, J. High Energy Phys. 11 (2012) 110.
\bibitem{Wu2}C. Niu, Y. Tian, X. N. Wu, \PRD 85, 024017 (2012).
\bibitem{LYX}S. W. Wei, Y. X. Liu, \PRD 87, 044014 (2013).
\bibitem{Ma1}R. Zhao, H. H. Zhao, M. S. Ma, L. C. Zhang, Eur. Phys. J. C 73, 2645(2013) .
\bibitem{Ma3}M. S. Ma, Y. Q. Ma, \CQG,  32 (2015) 035024.
\bibitem{ZhaoHH.2015}H. H. Zhao, L. C. Zhang, M. S. Ma and R. Zhao, \CQG 32 (2015) 145007.
\bibitem{Pa1}T. Padmanabhan, arXiv:0202080.
\bibitem{Pa2}T. Padmanabhan, \CQG 19(2002) 5387.
\bibitem{Pad:2006}A. Paranjape, S. Sarkar and T. Padmanabhan, \PRD 74,104015(2006).
\bibitem{Cai:2010}R. G. Cai, N. Ohta, \PRD  ~{\bf 81}, 084061 (2010).
\bibitem{Ma2}M. S. Ma, R. Zhao, \PLB 751, 278 (2015).
\bibitem{Hansen}D. Hansen, D. Kubiz\v{n}\'{a}k, R. B. Mann, \JHEP 01(2017)047 , arXiv:1603.05689.
\bibitem{Horava}P. Ho\v{r}ava, \PRD ~{\bf 79}, 084008 (2009).
\bibitem{Lu:2009}H. Lu, J. W. Mei, C. N. Pope, \PRL {\bf 103}, 091301 (2009).
\bibitem{Cai:2009-PRD}R. G. Cai, L. M Cao, N. Ohta, \PRD ~{\bf 80}, 024003 (2009).
\bibitem{Cai:2009-PLB}R. G. Cai, L. M Cao, N. Ohta, Physics Letters B {\bf 679}, 504 (2009).
\bibitem{EK:2010}E. Kiritsis, G. Kofinas, JHEP 01, 122(2010).
\bibitem{Iran}J. Sadeghi, K. Jafarzade, B. Pourhassan, \IJTP 51,3891 (2012).
\bibitem{Lu:2014}Hai-Shan Liu, Hong Lu, JHEP12, 071(2014).

\bibitem{Chen:2011}Q. J. Cao, Y. X. Chen, K. N. Shao, \PRD  ~{\bf 83}, 064015 (2011).
\bibitem{Majhi:2012}B. R. Majhi, D. Roychowdhury, Class. Quantum Grav. {\bf 29}, 245012 (2012).
\bibitem{Mo}J. X. Mo, X. X. Zeng, G. Q. Li, X. Jing, W. B. Liu, \JHEP 10, 056(2013).
\bibitem{Callen}H. B. Callen, `` Thermodynamics and an Introduction to Thermostatistics", John Wiley \& Sons, New York, NY, USA,1985.
\bibitem{Davies}P. Davies, Proc. R. Soc. A 353 499(1977).
\bibitem{MoJ}J. X. Mo,  Astrophys Space Sci, 356, 319(2015).
\bibitem{FAM}A. M. Frassino, D. Kubiznak, R. B. Mann, F. Simovic, \JHEP 2014, 9(2014) 080.
\bibitem{Dolan2014} B. P. Dolan, A. Kostouki, D. Kubiznak and R. B. Mann, \CQG 31, 242001(2014).
\bibitem{NA}N. Altamirano, D. Kubiznak, R.B. Mann and Z. Sherkatghanad, Class. Quant. Grav. 31 (2014) 042001.
\bibitem{LYX2}S.-W. Wei and Y.-X. Liu, Phys. Rev. D 90 (2014) 044057.

















\end{thebibliography}
\end{document}